\documentclass[10pt]{iopart}

\usepackage{iopams}
\usepackage[T1]{fontenc}
\usepackage[utf8]{inputenc}
\usepackage{graphicx}
\usepackage{tabularx}
\usepackage{color}
\usepackage{textcomp}
\usepackage{enumerate}
\usepackage{todonotes}
\usepackage{hyperref}
\usepackage{comment}
\usepackage{url}
\usepackage{caption}
\usepackage{iopams}
\usepackage{cite}
\usepackage{url}
\usepackage{color,soul}
\usepackage{bm}
\usepackage{ulem}

\usepackage{xcolor}

\newcommand{\revv}[1]{{#1}}
\newcommand{\rev}[1]{{#1}}

\usepackage{todonotes}

\newcommand{\vecb}[1]{\boldsymbol{#1}}
\renewcommand{\vec}[1]{\vecb{#1}}
\newcommand{\hatb}[1]{\hat{\boldsymbol{#1}}}

\begin{document}

\title{3D simulations of positive streamers in air in a strong external magnetic field}
\author{Zhen Wang$^{1,2}$, Anbang Sun$^{2}$, Sa\v{s}a Dujko$^3$, Ute Ebert$^{1,4}$, Jannis Teunissen$^{1}$}

\address{$^1$Centrum Wiskunde \& Informatica, Amsterdam, The Netherlands\\
  $^2$State Key Laboratory of Electrical Insulation and Power Equipment, School of Electrical Engineering, Xi'an Jiaotong University, Xi'an, 710049, China\\
  $^3$Institute of Physics, University of Belgrade, Serbia\\
  $^4$Dept. Applied Physics, Eindhoven University of Technology, Eindhoven, The Netherlands 
}
\ead{jannis.teunissen@cwi.nl}

\begin{abstract}
  We study how external magnetic fields from 0 to 40~T influence positive streamers in atmospheric \rev{pressure} air, using 3D PIC-MCC (particle-in-cell, Monte Carlo collision) simulations.
  When a magnetic field $\vec{B}$ is applied perpendicular to the background \rev{electric} field $\vec{E}$, the streamers \rev{deflect towards the $+\vec{B}$ and $-\vec{B}$ directions which results in a branching into two main channels.}
  With a stronger magnetic field the \rev{angle between the branches} increases, and for the 40~T case the branches grow almost parallel to the magnetic field.
  \rev{Due to the $\vec{E}\times\vec{B}$ drift of electrons we also observe a streamer deviation in the opposite $-\vec{E}\times\vec{B}$ direction, where the minus sign appears because positive streamers propagate opposite to the electron drift velocity.
    The deviation due to this $\vec{E}\times\vec{B}$ effect is smaller than the deviation parallel to $\vec{B}$.}
  \rev{In both cases of $\vec{B}$ perpendicular and parallel to $\vec{E}$,} the streamer radius decreases with the magnetic field strength.
  \rev{We relate our observations to the effects of electric and magnetic fields on electron transport and reaction coefficients.}
\end{abstract}

\vspace{2cm}

\maketitle

\ioptwocol

\section{Introduction}

Streamer discharges are often the first stage in the electric breakdown of gases~\cite{nijdam2020physics}.
They are ionized channels that rapidly grow due to strong enhancement of the electric field at their tips; this high local field causes electron impact ionization which lets the plasma channel grow.
In air, the growth of positive streamers against the electron drift direction is supported by nonlocal photoionization near regions of high impact ionization. 
The ionized paths created by streamers can later turn into sparks and lightning leaders, through Ohmic heating and gas expansion.
A streamer corona paves the way of lightning leaders, and streamers are directly visible as huge sprite discharges below the ionosphere~\cite{pasko2006theoretical,ebert2010review}.
Streamers also play a role in technological applications, such as plasma medicine~\cite{keidar2013cold}, and plasma assisted combustion~\cite{starikovskaia2014plasma}.


Magnetic fields play an important role for many types of discharges and plasmas, but for streamer discharges magnetic effects are usually not considered. 
The magnetization of electrons can be expressed by the Hall parameter $\beta_\mathrm{Hall} = \omega_\mathrm{ce}/\nu$, where $\omega_\mathrm{ce} = eB/m_e$ is the electron gyrofrequency in the magnetic field strength $B$, $\nu$ is the electron-neutral collision frequency, \rev{$e$ is the elementary charge and $m_e$ is the electron mass}, see e.g.~\cite{Hagelaar_2007}.
Electrons contributing to the growth of a streamer discharge typically have a high energy (of multiple eV), and therefore a high collision frequency.
For example, in air at standard conditions the \rev{electron collision frequency in an electric field of $3 \, \textrm{MV/m}$ (which is approximately the breakdown field) is $\nu \sim 3 \times 10^{12} \, \mathrm{s}^{-1}$.
  A substantial effect of a magnetic field can be expected when $\beta_\mathrm{Hall} \sim 1$, see e.g.~\cite{ebert2010review,starikovskiy2021streamer}, which would require $B \sim 17 \, \mathrm{T}$.}
The magnetic fields induced by the currents inside a streamer are generally many order of magnitude weaker, as discussed in section 5.1 of~\cite{nijdam2020physics}.
Significant magnetic effects can therefore only come from an external magnetic field.

\rev{We remark that an estimate for the maximal magnetic field strength $B_\mathrm{max}$ induced by a streamer is given in section 5.1 of~\cite{nijdam2020physics}:
  \begin{equation}
    \label{eq:Bmax}
    B_\mathrm{max} \approx v E_\mathrm{max} / c^2,
  \end{equation}
  where $v$ is the streamer velocity, $E_\mathrm{max}$ the maximal electric field at the streamer head and $c$ the speed of light.
This approximation is valid regardless of the gas number density.
  }

Since $\nu$ is proportional to the gas number density $N$ \rev{(with small corrections due to three-body processes)}, the Hall parameter depends on the reduced magnetic field $B/N$. 
Ness proposed the Huxley as a unit of $B/N$ (1 Hx = $10^{-27}~\mathrm{T\, m^3}$) that is commonly utilized in swarm studies of electron transport in electric and magnetic fields~\cite{ness1994multi}.
This scaling with gas density is similar to the dependence of \rev{(reduced) electron transport coefficients} on the reduced electric field $E/N$~\cite{pasko2007red,ebert2010review}. A streamer at ground pressure in an electric field of 1.5~MV/m and in a magnetic field of 10~T therefore scales approximately to a sprite streamer at 83~km altitude, hence in an air density of $N=10^{-5}N_{\rm ground}$, in an electric field of 15~V/m and in a magnetic field of 100~$\mu$T.
\rev{As discussed in~\cite{Pasko_1997,ebert2010review}, the geomagnetic field at the equator is weaker (about 30~$\mu$T), which is why sprites on earth are not seen to be magnetized.} (The geomagnetic field rises to about 60 $\mu$T near the poles at sprite altitude, but there the field direction is vertical, and lightning and sprites are rare.)
  However, on Jupiter streamer discharges in sprites and lightning could be magnetized in the strong and irregular magnetic field of that planet~\cite{connerneyNewModelJupiter2022, bloxhamDifferentialRotationJupiter2022, kolmasovaLightningJupiterPulsates2023, gilesPossibleTransientLuminous2020}. 

There have only been a few experimental studies on streamers in magnetic fields, as it is challenging to obtain a strong enough magnetic field in a sufficiently large volume.
In~\cite{manders2008propagation} the early stages of magnetized streamers were studied in $99.9\%$-pure nitrogen at pressures of 0.27 to 0.8\,bar, using a magnetic field strength of up to 12.5\,T.
For negative streamers a clear bending in the $\vec{E} \times \vec{B}$ direction was observed, as expected for electrons drifting in a crossed electric field $\vec{E}$ and magnetic field $\vec{B}$, but for positive streamers, the experimental results were more difficult to interpret.
In earlier work, the effect of a magnetic field on surface discharges has also been studied using the Lichtenberg technique~\cite{uhlig1989spatial,hara1992deflection}.
In these studies, a clear bending of negative discharges was also observed, whereas positive streamers showed a smaller deviation.


Recently, the effect of an external magnetic field on streamer discharges has been investigated in two computational studies~\cite{starikovskiy2021streamer,Janalizadeh_2023} in which the magnetic field was assumed to be parallel to the background electric field.
In~\cite{starikovskiy2021streamer}, a 2D axisymmetric model was used to simulate both positive and negative streamers in an external parallel magnetic field. A decrease of streamer radius was observed for both streamer polarities. The authors attributed this `self-focusing' phenomenon to a sharp slowdown in the radial growth of the streamers.
The same phenomenon was recently also observed in~\cite{Janalizadeh_2023}, in which the effects of Jupiter's strong magnetic field (from
0.2 to 1.5 {\rm mT}) on streamer inception and propagation were studied.




In this paper, we generalize the above computational studies by also considering perpendicular magnetic fields, using 3D particle-in-cell simulations.
In order to explain the main propagation phenomena of streamers, electron transport data in electric and magnetic fields crossed at arbitrary angle are also \rev{presented}.
We focus on positive streamers in air at 1\,bar, and on magnetic field strengths of up to 40\,Tesla.
However, our results can be scaled to different pressures and corresponding field strengths, as discussed above.



\section{Model description}
\subsection{3D PIC-MCC model}

We use a PIC-MCC (particle-in-cell, Monte-Carlo Collision) model, which combines the particle model described in~\cite{teunissen20163d} with the Afivo AMR (adaptive mesh refinement) framework described in \cite{teunissen2018afivo}.
In this model, only free electrons are tracked as particles, ions are tracked as densities, and neutral gas molecules are included as a background that electrons stochastically collide with.
We use Phelps' cross sections for N$_2$ and O$_2$ \cite{phelps1985anisotropic,Phelps-database}.
To use them in particle simulations, we assume isotropic electron scattering and convert the effective momentum transfer cross-sections to elastic momentum transfer cross sections by subtracting the sum of the
inelastic cross sections~\cite{Pitchford_1982}.



An advantage of a PIC-MCC model is that a magnetic field can relatively easily be included by modifying the particle mover, see section~\ref{sec:particle-mover}.
To include a magnetic field in a fluid model is more complicated, since both the computation of transport data and the inclusion of such data into the model are non-trivial, see e.g.~\cite{Dujko_2010,janalizadeh2023efficient}.

\subsubsection{Photoionization. $\;$}

\revv{Zheleznyak's} photoionization model~\cite{Zheleznyak_1982} is included as a stochastic process, as described in~\cite{chanrion2008pic,teunissen20163d}.
We briefly summarize the Monte Carlo method below.
When a simulated electron with a weight $w$ ionizes a neutral molecule, the number of ionizing photons that is generated is sampled from the Poisson distribution with mean
\begin{equation}
  n_\mathrm{photons} = \frac{p_q}{p+p_q}\xi w,
  \label{eq:UV-source-term}
\end{equation}
where $p$ is the gas pressure, $p_q = 40$ mbar is the quenching pressure, and $\xi = 0.075$ is a proportionality factor that we assume to be constant for simplicity.
Note that the photons thus have a weight of one.
For each photon, an absorption length is sampled from the absorption function
\begin{equation}
    f(r) = \frac{{\exp ( - {\chi _{\min }}{p_{{o_2}}}r) - \exp ( - {\chi _{\max }}{p_{{o_2}}}r)}}{{r\ln ({\chi _{\max }}/{\chi _{\min }})}},
    \label{eq:absorption-fuction}
  \end{equation}
  as described in~\cite{chanrion2008pic,teunissen20163d}.
  Here $\chi_{max} = 1.5 \times 10^2/(\textrm{mm bar})$, $\chi_{min} = 2.6/(\textrm{mm bar})$, and $p_{O_2}$ is the partial pressure of oxygen.
  An isotropic direction is then sampled, after which an ionization event of O$_2$ is generated at the location of absorption if this location is inside the gas region of the computational domain.

\subsubsection{Super-particles. $\;$}
\label{sec:superparticles}

\rev{So-called super-particles~\cite{Hockney_1988} are used to speed up the simulations and save memory.
  The weight parameter $w_i$ determines how many physical particles the $i$\textsuperscript{th} simulation particle represents.
  During a simulation, the weights $w_i$ change over time by merging and splitting particles as described in~\cite{Wang_2022}.
  Particle weights are updated when the number of simulation particles has grown by a factor of $1.25$ or following a change of the AMR mesh (see section \ref{sec:AMR}), so that they stay close to a desired weight $w$ given by}
\begin{eqnarray}
  \rev{w} =  n_e\times \Delta V / N_{ppc},
  \label{eq:weight-calculation-superparticle}
\end{eqnarray}
\rev{where $n_e$ is the electron density in a cell, $\Delta V$ the cell volume and $N_{ppc}$ is the target number of simulation particles per cell, here set to $N_{ppc} = 75$.}

\subsubsection{Particle mover. $\;$}
\label{sec:particle-mover}

We use Boris' rotation method~\cite{Birdsall_2004} to advance the position and velocity of electrons in time.
The timestep in our simulations is limited by several restrictions
\begin{equation}
    \Delta{t} \leq \mathrm{min}(0.5 \times \Delta x_\mathrm{min} / \tilde{v}_{\mathrm{max}},\Delta {t_\mathrm{drt}}, \revv{0.63 \times \omega_\mathrm{ce}}).
  \end{equation}
  Here $\Delta x_{\mathrm{min}}$ indicates the minimal grid spacing, and $\tilde{v}_{\mathrm{max}}$ is an estimate of the particle velocity at the 90\%-quantile.
  $\Delta t_\mathrm{drt}$ is the Maxwell time, also known as the dielectric relaxation time, which is a typical time scale for electric screening.
  Finally, the last criterion ensures that the gyration of electrons is accurately resolved.

\subsubsection{Adaptive mesh refinement. $\;$}
\label{sec:AMR}
Adaptive mesh refinement (AMR) is used for both computational efficiency and computational accuracy.
\rev{The mesh is refined} based on the following criteria~\cite{teunissen2017simulating}:
\begin{itemize}
\item refine if $\alpha(E) \Delta x > 1.0$,
\item de-refine if $\alpha (E) \Delta x < 0.125$, but only if $\Delta x$ is smaller than 10 $\mu$m.
\end{itemize}
Here $\alpha(E)$ is the field-dependent ionization coefficient, and $\Delta x$ is the grid spacing, which is bound by $\rev{2\,\mu\textrm{m}}\leq \Delta x \leq 0.4~\textrm{mm}$.

\subsection{Computational domain and simulation conditions}
\label{sec:initial-condition}

\begin{figure}
  \centering
  \includegraphics[width=\linewidth]{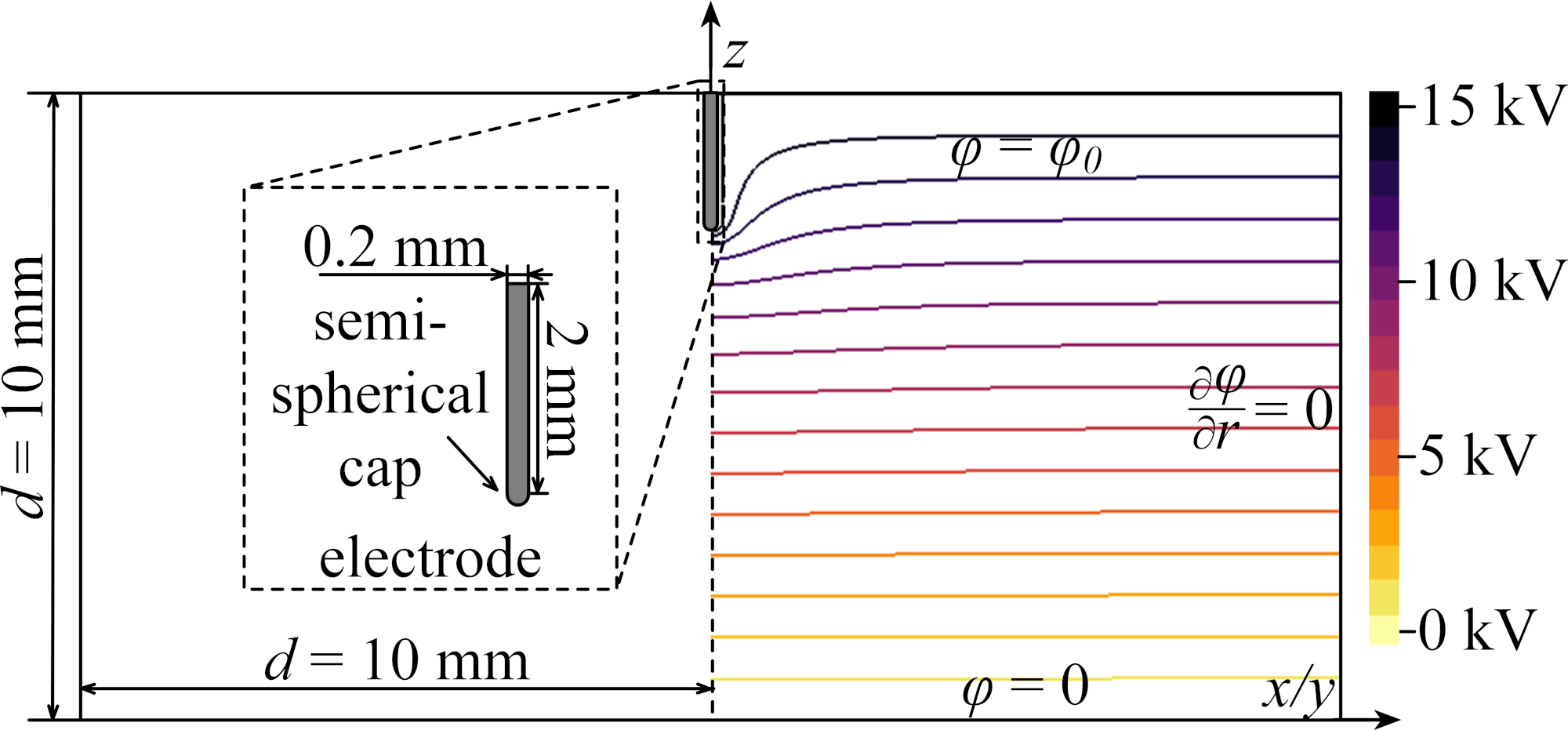}
  \caption{Cross section of the 3D computational domain, which measures 20\,mm $\times$ 20\,mm $\times$ 10\,mm.
    \rev{The electric potential $\varphi$ (in the absence of a discharge) and the electrode geometry are shown.
      A voltage of $\varphi_0 = 15 \, \mathrm{kV}$ is applied at the top of the domain and the needle electrode.
      The bottom of the domain is grounded ($\varphi = 0$), and at the sides a Neumann zero boundary condition is used for $\varphi$.}}
  \label{fig:initial-condition}
\end{figure}

Simulations are performed in artificial air, containing 80$\%$ N$_2$ and 20$\%$ O$_2$, at $p$ = 1 bar and $T$ = 300 K.
Figure~\ref{fig:initial-condition} shows a cross section of the 20\,mm $\times$ 20\,mm $\times$ 10\,mm computational domain.
A rod-shaped electrode with a semi-spherical cap is placed at the center of the domain~\cite{Teunissen_2023}.
This electrode is 2\,mm long and has a radius of 0.2\,mm.
\rev{Boundary conditions for the electric potential are given in the caption of figure~\ref{fig:initial-condition}.}
\rev{In our computational domain, the background electric field points in the $-z$ direction, and a magnetic field is applied in either the $-z$ direction (parallel case) or in the $-x$ direction (perpendicular case).}

There is initially no background ionization besides an electrically neutral plasma seed, which is placed at the tip of the electrode to provide initial ionization.
Electrons and positive ions are generated by sampling from a Gaussian distribution
\begin{equation}
  n_i(\mathbf{r})=n_e(\mathbf{r})= 10^{16}\,\mathrm{m}^{-3}
  \exp\left[\frac{-\rev{|}\mathbf{r}-\mathbf{r_0}\rev{|}^2}{(0.1 \, \mathrm{mm})^2}\right],
  \label{eq:gaussian-distribution}
\end{equation}
where $\mathbf{r_0}$ is the location of the tip of the electrode, \rev{given by $(x, y, z) = (10, 10, 7.8) \, \textrm{mm}$}.

\subsection{\rev{Effect of magnetic field on electron drift and ionization}}
\label{sec:transport-data}

\begin{figure*}
  \centering
  \includegraphics[width=\linewidth]{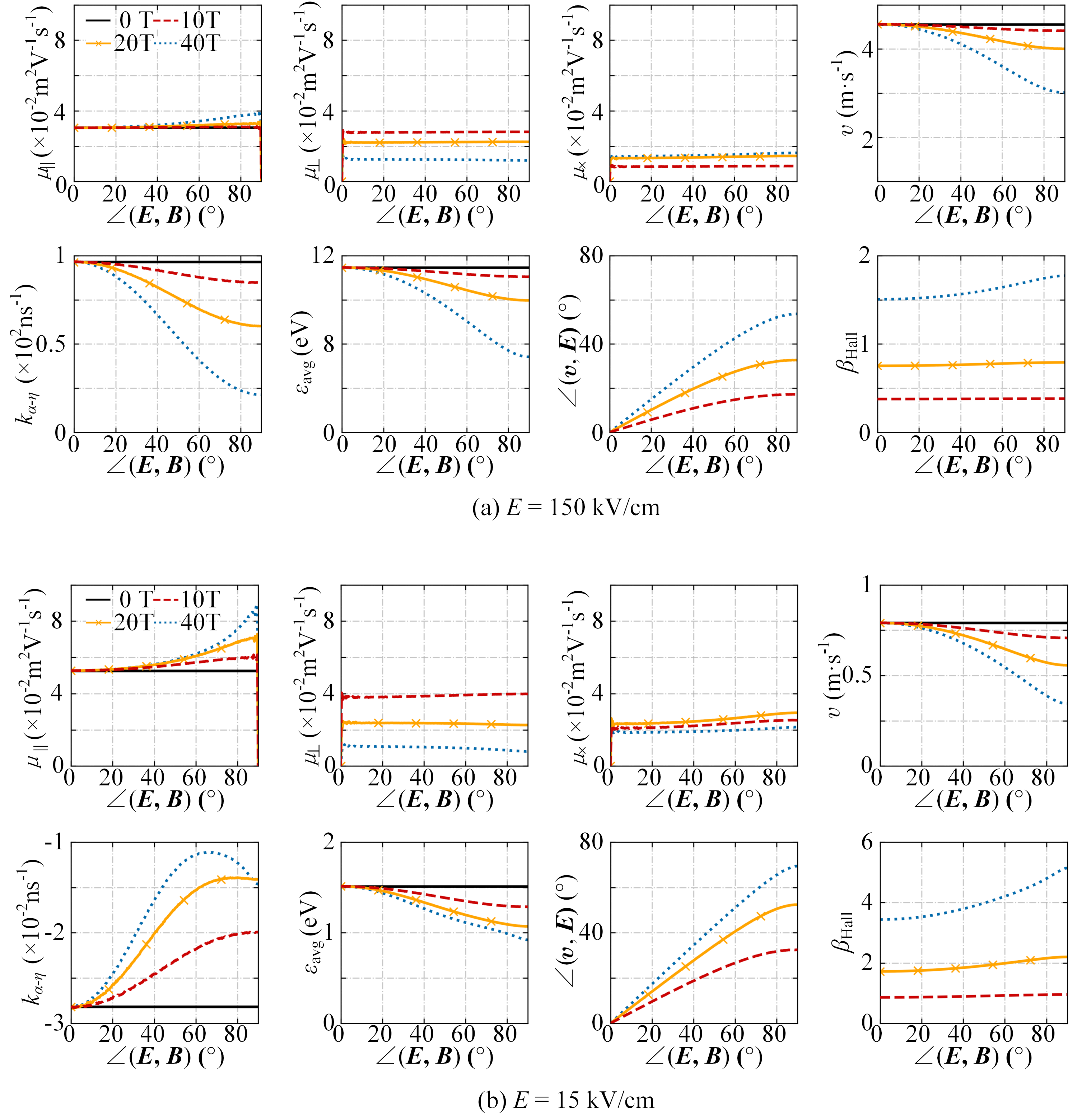}
  \caption{Electron transport coefficients in homogeneous $\vec{E}$ and $\vec{B}$ fields as a function of the angle $\angle(\vec{E},\vec{B})$ between the fields.
    The data was generated (a) for a typical maximal field at the streamer head of $E = 150 \, \mathrm{kV/cm}$ and (b) for half the breakdown field $E = 15\,\mathrm{kV/cm}$, and for magnetic fields of $B = 0, 10, 20, 40\,\mathrm{T}$, in synthetic air at 1 bar and 300 K.
    Here $\mu_\parallel$ and $\mu_\bot$ are the flux mobility components parallel and perpendicular to $\vec{B}$, and $\mu_{\times}$ is a flux mobility component in the $E \times B$ direction, see section~\ref{sec:transport-data}.
    $v$ is the electron drift velocity, $k_{\alpha-\eta}$ is the effective ionization rate (positive for $E = 150 \, \mathrm{kV/cm}$ and negative for $E = 15\,\mathrm{kV/cm}$), and $\varepsilon_\mathrm{avg}$ is the mean electron energy.
    $\angle (\vec{v},\vec{E})$ is the angle between the electron drift velocity $\vec{v}$ and $\vec{E}$, and $\beta_\mathrm{Hall}$ is the Hall parameter \rev{for the ensemble of electrons, see section~\ref{sec:transport-data}}.}
  \label{fig:drift_all}
\end{figure*}

\rev{To understand the behavior of streamer discharges in $\vec{E}$ and $\vec{B}$ fields, it helps to know how free electrons behave in these fields.
We have therefore} computed electron transport coefficients for homogeneous $\vec{E}$ and $\vec{B}$ fields at an arbitrary angle
with a Monte Carlo swarm code \url{https://github.com/MD-CWI/particle_swarm}.
Data computed for background electric fields of $150~\mathrm{kV/cm}$ and $15~\mathrm{kV/cm}$ and for magnetic fields between 0 and 40\,T are shown in figure~\ref{fig:drift_all} \rev{as a function of the angle between $\vec{E}$ and $\vec{B}$}. The Hall parameter $\beta_\mathrm{Hall}$ \rev{for the ensemble of electrons} is also indicated. \rev{In this case, we define $\beta_\mathrm{Hall} = \omega_\mathrm{ce}/\bar{\nu}$ where $\bar{\nu}$ is the average electron collision frequency~\cite{Janalizadeh_2023}.}

For the analysis of electron motion, we decompose the electric field into a part parallel and perpendicular to $\vec{B}$ as $\vec{E} = \vec{E}_{\parallel} + \vec{E}_{\bot}$.
Then three components of the electron mobility can be distinguished: 
\begin{eqnarray}
  \label{eq:mobilities-par}
  \mu_{\parallel} = v_{\parallel}/E_{\parallel} = |\vec{v} \cdot \hatb{E}_{\parallel}|/E_{\parallel},\\
  \mu_{\bot} = v_{\bot}/E_{\bot} = |\vec{v} \cdot \hatb{E}_{\bot}|/E_{\bot},\\
  \label{eq:mobilities-cross}
  \mu_{\times} = v_{\times}/E_{\bot} = |\vec{v} \cdot (\rev{\hatb{E}_{\bot}} \times \hatb{B})|/E_{\bot},
\end{eqnarray}
where $\vec{v}$ is the electron drift velocity, and $\mu_{\parallel}$, $\mu_{\bot}$ and $\mu_{\times}$ are respectively the mobility parallel to the magnetic field, the mobility perpendicular to the magnetic field \rev{(but parallel to $\vec{E}_{\bot}$)}, and the ``mobility'' in the $\vec{E} \times \vec{B}$ direction. 
Here $\hatb{E}$ and $\hatb{B}$ denote unit vectors in the direction of the $\vec{E}$ and $\vec{B}$ respectively.
\rev{Note that $E_{\parallel} = E \cos(\theta)$ and $E_{\bot} = E \sin(\theta)$, where $\theta = \angle(\vec{E}, \vec{B})$ is the angle between $\vec{E}$ and $\vec{B}$.
Furthermore we remark that equations (\ref{eq:mobilities-par}--\ref{eq:mobilities-cross}) define so-called flux mobilities, see e.g.~\cite{Petrovic_2009}.}


A clear effect of a stronger magnetic field is that \rev{$\mu_{\bot}$} is reduced.
The reduction in $\mu_{\bot}$ leads to a lower mean electron energy and a smaller ionization coefficient when the angle between $\vec{E}$ and $\vec{B}$ increases, \rev{because the energy electrons on average gain per unit time from the electric field is given by $e\mu_{\parallel} E_{\parallel}^2 + e\mu_{\bot} E_{\bot}^2$}.
For $\vec{B} = 40~\mathrm{T}$, the ionization rate is reduced by almost $80\%$ when $\vec{E}$ and $\vec{B}$ are perpendicular.
The reduction in mean electron energy \rev{is also related to} an increase in the parallel electron mobility $\mu_{\parallel}$, since electron mobilities are \rev{typically} higher at lower electron energies.

The magnitude of the $\vec{E} \times \vec{B}$ drift, here denoted by
\begin{equation}
  \rev{v_\times = \mu_{\times} E_{\bot} = \mu_{\times} E \sin(\theta),}
\end{equation}
depends on the magnetization of electrons and on the respective fields.
In the absence of collisions
\begin{equation}
  \rev{v_\times = |\vec{E} \times \vec{B}|/B^2 = E \sin(\theta)/B,}
\end{equation}
so that a stronger magnetic field leads to a smaller $v_\times$.
However, when collisions are included a lower magnetic field will lead to a lower magnetization (i.e., Hall parameter $\beta_\mathrm{Hall}$), with $v_\times \to 0$ for $\beta_\mathrm{Hall} \to 0$.
For a given electric field, $v_\times$ will thus first increase with the magnetic field \rev{strength} and then decrease, as can be seen \rev{from the $\mu_\times$ plot in figure~\ref{fig:drift_all}b}.
\revv{More specifically, it can be shown that to a good approximation $\mu_{\times} = \mu_{\parallel} \, \beta_\mathrm{Hall} / (1 + \beta_\mathrm{Hall}^2)$~\cite{janalizadeh2023efficient}, which has a maximum at $\beta_\mathrm{Hall} = 1$.}

With a magnetic field the mobilities are $\mu_{\bot} < \mu_{\parallel}$ and $\mu_{\times} > 0$, which means that the electron drift velocity $\vec{v}$ makes an angle with the electric field $\vec{E}$.
This angle can be as large as $70^\circ$ for the conditions considered here, as shown in figure~\ref{fig:drift_all}.

Note that when the electric and the magnetic field are parallel electron transport and reaction coefficients (except for transverse diffusion coefficients) hardly depend on the magnetic field strength.
In this case, the magnetic field does not affect the energy gain of electrons due to their acceleration parallel to $\vec{E}$.



\section{Results \& Discussion}
\label{subsec:results}

\subsection{3D simulations}

We simulate positive streamers with an external magnetic field of 0, 10, 20 and 40~T.
In these simulations, the background electric field $E = 15~\textrm{kV/cm}$ \rev{between the plate electrodes} is about half of the breakdown field, \rev{and it} points downwards in $-z$ direction, see Fig.~\ref{fig:initial-condition}. 
The magnetic field $B$ either points in the same direction (parallel case) or in \rev{the $-x$} direction (perpendicular case).
\rev{In all cases, the positive streamers grow due to photoionization which produces free electrons ahead of them~\cite{nijdam2020physics}.}

\begin{figure}
  \centering
  \includegraphics[width=\linewidth]{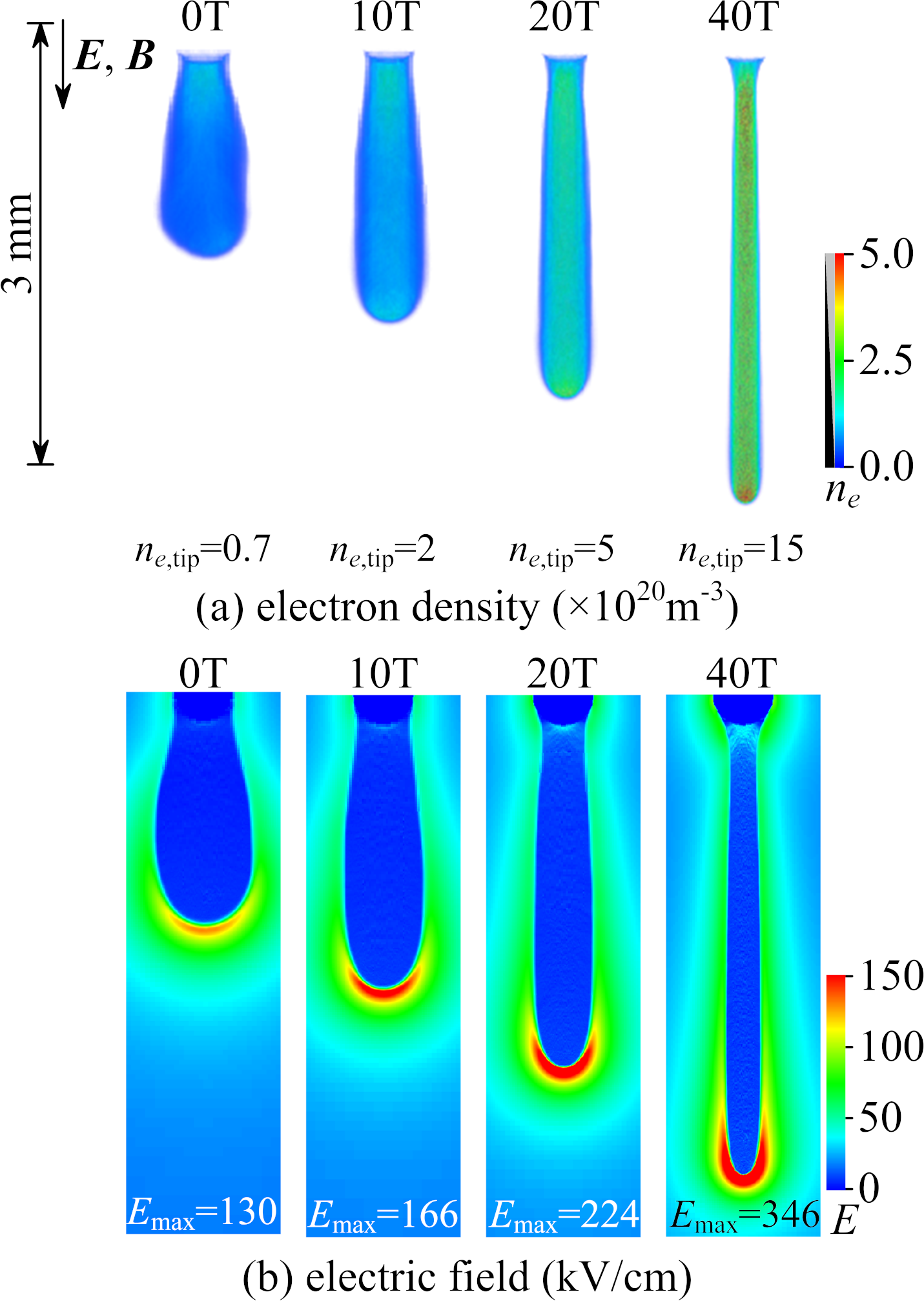}
  \caption{Simulations of positive streamers at $t = 3~\mathrm{ns}$ in magnetic fields of 0, 10, 20 and 40~T that are parallel to a background electric field of $15\mathrm{kV/cm}$. (a) volume rendering of the electron density, (b) cross section of the electric field. \rev{For reference, the electron density behind the streamer tip ($n_{e,\mathrm{tip}}$) and the maximal electric field strength $E_\mathrm{max}$ are indicated in the panels.}}
  \label{fig:par-comparison}
\end{figure}

Results with a parallel magnetic field are shown in figure~\ref{fig:par-comparison}. \rev{In agreement with previous work~\cite{starikovskiy2021streamer,Janalizadeh_2023}, the following main phenomena are observed:}
\begin{itemize}
  \item For a stronger magnetic field, the streamer diameter decreases and its velocity and the \rev{maximum} electron density inside the channel increase. 
  \item The streamer overall stays axisymmetric, if it started like this. 
\end{itemize}

\begin{figure*}
  \centering
  \includegraphics[width=\linewidth]{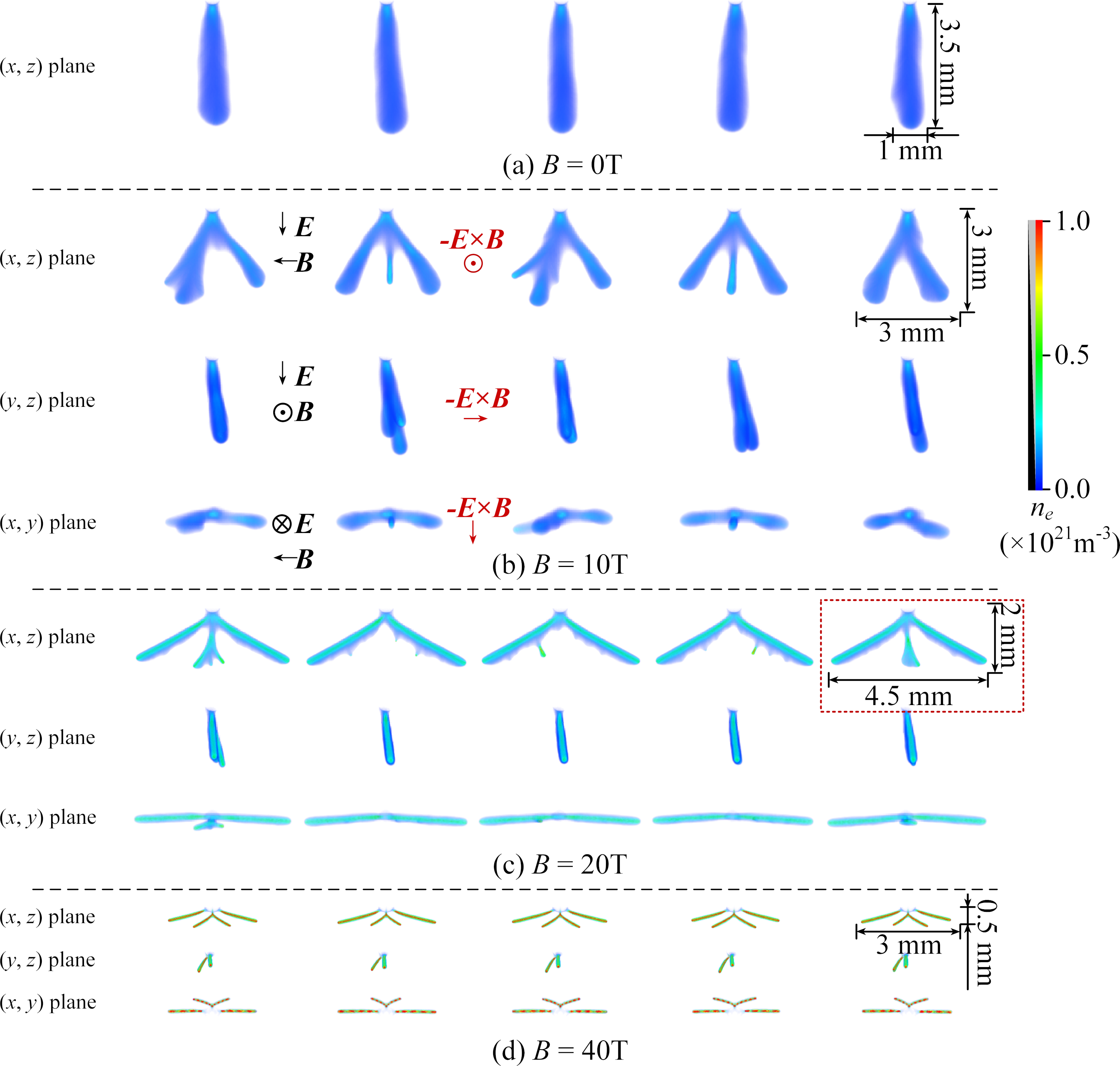}
  \caption{\rev{Simulations of positive streamers at $t = 6~\textrm{ns}$ in magnetic fields of 0, 10, 20 and 40~T that are perpendicular to a background electric field of $15\,\mathrm{kV/cm}$.
      Electron densities are visualized using 3D volume rendering.}
    Each row shows five runs in the same magnetic field to \rev{illustrate} the stochasticity of the particle simulations.
    \rev{For $B > 0$ several viewing angles are shown, with the plane that is viewed indicated on the left.
      The directions of $\vec{E}$, $\vec{B}$ and $-\vec{E} \times \vec{B}$ are indicated in the $B=10~\mathrm{T}$ case, they are the same for the $B=20~\mathrm{T}$ and $B=40~\mathrm{T}$ cases.}
    The simulation in the red dotted frame is shown in more detail in figure~\ref{fig:enlarged-field}.}

  \label{fig:ver-comparison}
\end{figure*}

\begin{figure}
  \centering
  \includegraphics[width=\linewidth]{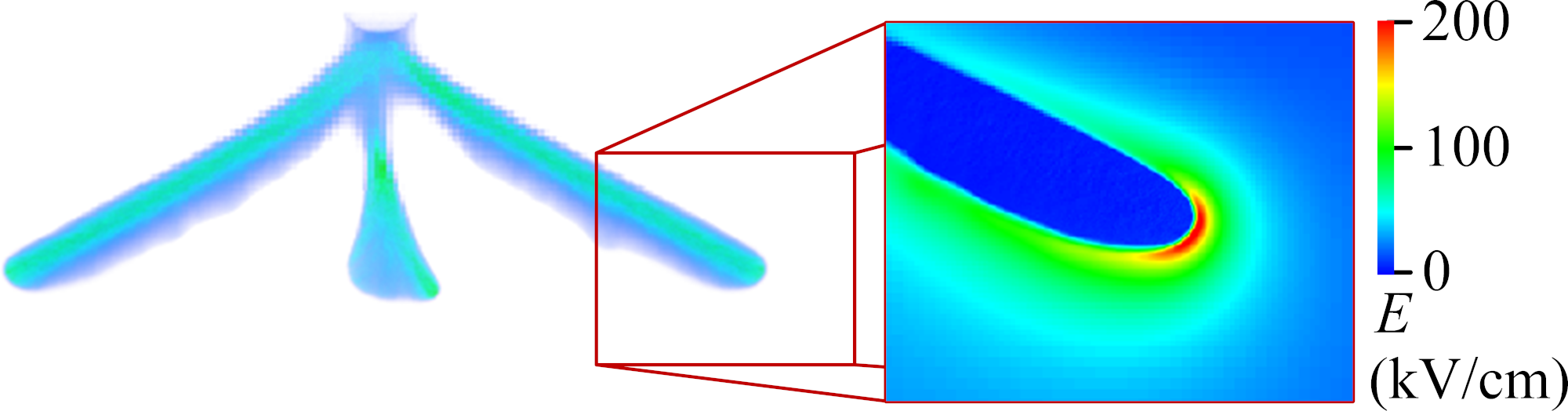}
  \caption{Left: Enlarged electron density in a magnetic field of $B = 20$~T. Shown is the rightmost case from Fig.~\ref{fig:ver-comparison} \rev{viewed in the $(x,z)$ plane} (marked with a red dotted frame in that figure). 
  Right: Further enlarged cross section of the electric field $E$.}
  \label{fig:enlarged-field}
\end{figure}

Results with a perpendicular magnetic field are shown in figure~\ref{fig:ver-comparison}, and one particular run in a field of 20~T is magnified in figure~\ref{fig:enlarged-field}. The main phenomena are: 
\begin{itemize}
  \item With growing magnetic field, the streamer diameter decreases and its velocity and the \rev{maximum} electron density inside the channel decrease as well. 
  \item \rev{The streamers deflect towards the $+\vec{B}$ and $-\vec{B}$ direction which results in a branching into two main channels.
    The angle between the branches increases with $B$.
    There seem to be two preferred streamer propagation directions in the plane spanned by $\vec{E}$ and $\vec{B}$.}
  \item The branched streamer does not completely lie in the plane spanned by $\vec{E}$ and $\vec{B}$, but shows a slight \rev{deviation towards} the $-\vec{E}\times \vec{B}$ direction.
\end{itemize}
We will explain these phenomena \rev{below, making use of the electron transport data presented in section~\ref{sec:transport-data}.}

\subsection{Branching in a perpendicular field}
\label{sec:branch-perp-field}

With a perpendicular $\vec{B}$-field of 20\,T to 40\,T positive streamers typically split into two main channels, which both lie approximately in the plane spanned by $\vec{E}$ and $\vec{B}$. The angle between the branches grows with the magnetic field, until the branches are almost parallel or antiparallel to $\vec{B}$.
The two main channels are quite symmetric in the plane spanned by $\vec{E}$ and $\vec{B}$, because changing $\vec{B}$ into $-\vec{B}$ only changes the chirality of the gyration about the magnetic field line; the electrons on the left branch therefore have chirality opposite to those on the right branch, but otherwise the same energy distributions, ionization rates etc.
With a perpendicular $\vec{B}$-field of only 10\,T, the branching appears to be more stochastic, with a smaller angle between the branches and sometimes a third branch.
\rev{We remark that stochastic streamer branching is a common phenomenon in streamer discharges without a magnetic field, in which stochastic fluctuations can trigger a Laplacian instability, see e.g.~\cite{Liu_2004a,Luque_2011c,wangQuantitativeModelingStreamer2023}.
The peculiar aspect of the branching observed here is that it is not (or hardly) stochastic, but that it is rather induced by the two preferred propagation directions.}

While the direction of the primary streamers in higher magnetic fields is clearly determined by magnetic and electric fields (and possibly by the streamer radius), \rev{the shorter secondary branches that form at a later time} propagate in an electric field modified by the primary streamers. Therefore they deviate from the direction of the primary steamers.

\rev{A perpendicular magnetic field can contribute to streamer branching in two ways.}
The first is related to the ionization rate.
Without a magnetic field, the ionization rate depends only on the electric field strength, which is highest in the forward direction.
However, with a magnetic field, the ionization rate will also depend on the angle with the magnetic field, as shown in figure \ref{fig:drift_all}.
Since the strongest reduction in the ionization rate will occur for the lowest electron energies, i.e., when the electric field is perpendicular to $\vec{B}$, the maximum of the ionization rate can then lie at some intermediate angle between the background field $\vec{E}$ and the magnetic field $\vec{B}$.

The second effect is on the screening inside the streamer channel.
Perpendicular to $\vec{B}$, electric screening is slowed down due to the lower drift velocity, see figure \ref{fig:drift_all}.
Since electron screening parallel to $\vec{B}$ is not affected (the parallel mobility can even increase, as shown in figure~\ref{fig:drift_all}), there will be stronger field enhancement parallel to $\vec{B}$.
Both the change in ionization rate and the change in drift velocity will deform the streamer head, as shown in Fig.~\ref{fig:enlarged-field}, and contribute to branching in the $(\vec{E}, \vec{B})$-plane.

Unlike `normal' streamer branching~\cite{wangQuantitativeModelingStreamer2023}, branching in a strong magnetic field is a rather deterministic process.
The angle at which branches grow depends on the `competition' between the magnetic field, which favors growth parallel or anti-parallel to it, and electric field enhancement, which is strongest parallel to the background electric field.
\rev{A stronger} magnetic field therefore leads to a larger branching angle.

\subsection{Effect of magnetic field on streamer radius}
\label{sec:effect-magn-field-radius}

Figure~\ref{fig:par-comparison} shows that a parallel $\vec{B}$-field leads to a smaller streamer radius, consistent with the findings of~\cite{starikovskiy2021streamer,Janalizadeh_2023}.
The underlying mechanism is a reduction in the radial growth of the streamer, which is perpendicular to the $\vec{B}$-field, due to a lower ionization rate, see figure~\ref{fig:drift_all}.
Since the forward growth is not affected, as it is parallel to the $\vec{B}$-field, the result is a smaller radius and stronger electric field enhancement.

Figure~\ref{fig:ver-comparison} shows that a perpendicular $\vec{B}$-field also leads to a reduction in streamer radius.
We think there are two main mechanisms that play a role here.
The first is that after the streamers branch, the magnetic field is partially aligned with their growth velocity.
This parallel component of the magnetic field will have a similar effect as in the case where the electric and magnetic field are initially parallel, i.e., it will reduce the streamer radius.
The second mechanism is that the ionization rate around the streamers is reduced, since there is also a magnetic field component perpendicular to the streamer velocity.
This reduced ionization rate will probably play a similar role as a weaker background field (or a lower applied voltage) does in cases without a magnetic field, namely the formation of thinner channels.

Note that although the effect of a magnetic field on the streamer radius is similar for the parallel and perpendicular configurations, the effect on the streamer velocity is different: a parallel magnetic field leads to a higher velocity, see figure~\ref{fig:par-comparison}, whereas a perpendicular magnetic field leads to a lower velocity, see figure~\ref{fig:ver-comparison}.

\subsection{Bending in $-E \times B$ direction}
\label{sec:bending-ExB}

With a perpendicular magnetic field of $10 \, \textrm{T}$ and $20 \, \textrm{T}$, the discharge channels bend slightly towards the $-\vec{E} \times \vec{B}$ direction, as shown in figure~\ref{fig:ver-comparison}.
At $10 \, \textrm{T}$ the bending angle is about $9^\circ$ and at $20 \, \textrm{T}$ it is about $7^\circ$.
This bending is due to the $\vec{E} \times \vec{B}$ drift of electrons, which leads to a deviation in the opposite direction ($-\vec{E} \times \vec{B}$) since positive streamers propagate in the opposite direction of the electron drift velocity.


In figure~\ref{fig:drift_all} the angle the electron drift velocity makes with respect to $\vec{E}$ is shown.
When $\vec{E}$ and $\vec{B}$ are perpendicular, this angle ranges from about $18^\circ$ ($10 \, \textrm{T}$, $150 \, \textrm{kV/cm}$) to $70^\circ$ ($40 \, \textrm{T}$, $15 \, \textrm{kV/cm}$), due to the $\vec{E} \times \vec{B}$ drift.
These angles are considerably larger than the positive streamers' bending angle of up to about $10^\circ$, especially when comparing against the transport data in a lower background field of $15 \, \textrm{kV/cm}$.
We think there are several reasons for this. 
\rev{First of all, it should be noted that although the background electric field is perpendicular to $\vec{B}$ in the simulations, the enhanced electric field near the streamer head will generally not be perpendicular to $\vec{B}$.
Furthermore, streamers} grow more parallel to $\vec{B}$ as the magnetic field strength is increased, which reduces the magnitude of the $\vec{E} \times \vec{B}$ drift and thus also the bending angle.
This could explain why the observed bending is a bit smaller at 20\,T than at 10\,T, and why no clear bending can be observed at $40 \, \textrm{T}$.
A second reason is that the bending probably originates from the deformation of the streamer head, where the increased electron drift velocity increases electric field enhancement in the $-\vec{E} \times \vec{B}$ direction.
The high electric field at the streamer head will result in a smaller bending angle, as illustrated by figure~\ref{fig:drift_all}.
A third reason is that streamer growth depends on both the electron drift and the impact ionization rate (which depends on the electron energy distribution).
  There is only an $\vec{E} \times \vec{B}$ effect on the electron drift, in contrast to the branching mechanism discussed in section~\ref{sec:branch-perp-field}, where both the electron drift and the ionization rate depended on the angle with the magnetic field.

Finally, we remark that a side branch in the $\vec{E} \times \vec{B}$ direction is visible in all cases at $40 \, \textrm{T}$, which is probably caused by the effect the $\vec{E} \times \vec{B}$ drift has on the initial seed.

\subsection{Comparison with experimental work}
\label{sec:experimenta-comparison}

As mentioned in the introduction, the effect of a strong magnetic field on positive and negative streamers in nitrogen has been experimentally studied in~\cite{manders2008propagation}.
The discharges were observed in a plane perpendicular to $\vec{B}$, which means that only path deviations perpendicular to $\vec{B}$ could be observed, and thus not the branching phenomenon found here in the $\vec{E}, \vec{B}$-plane.
Negative streamers were found to clearly bend in the $\vec{E} \times \vec{B}$ direction, and a rather small deviation of positive streamers in the $-\vec{E} \times \vec{B}$ was observed.
We remark that after switching the voltage polarity, the $\vec{E} \times \vec{B}$ direction also flips, which means that in the experimental figures positive and negative streamers deviate in the same visual direction.

Earlier work used the Lichtenberg technique to study the effect of a magnetic field on surface discharges, with the magnetic field perpendicular to the surface.
In~\cite{uhlig1989spatial}, only negative discharges were observed, which showed a clear bending in the $\vec{E} \times \vec{B}$ direction.
In~\cite{hara1992deflection}, both polarities were considered, and it was found that negative streamers had a significantly larger deflection angle than positive ones.
As in~\cite{manders2008propagation}, streamers with both polarities bended in the same visual direction, which means that negative streamers deviated in the $\vec{E} \times \vec{B}$ and positive ones in the $-\vec{E} \times \vec{B}$ direction, consistent with our findings.

\section{Conclusions}
\label{sec:conclusions}

We have simulated the propagation of positive streamers in atmospheric air in external magnetic fields ranging from 0 to 40~T using a 3D PIC-MCC model including photoionization.
For magnetic fields perpendicular to the background electric field, \rev{streamers deflect towards the $+\vec{B}$ and $-\vec{B}$ direction resulting in a branching into two main channels.}
The angle between these branches increases with the magnetic field strength, and at $40 \, \textrm{T}$ they propagate almost parallel to the magnetic field, and thus almost perpendicular to the background \rev{electric} field.
We think there are two mechanisms that deform the streamer head and thereby contribute to this branching: the dependence of the ionization rate on the angle between the $\vec{E}$ and $\vec{B}$, and a reduction in electron drift velocity perpendicular to $\vec{B}$.
In agreement with earlier experimental work~\cite{uhlig1989spatial,hara1992deflection,manders2008propagation}, we observe that positive streamer slightly bend towards the $-\vec{E}\times \vec{B}$ direction, due to the $\vec{E}\times \vec{B}$ drift of electrons.
We also show that a perpendicular magnetic field reduces the streamer radius, a phenomenon that was earlier observed in axisymmetric simulations with a parallel magnetic field~\cite{starikovskiy2021streamer,Janalizadeh_2023}.
However, a difference between the parallel and perpendicular cases is that the streamer velocity increases with a parallel magnetic field whereas it decreases with a perpendicular magnetic field.

Finally, we \rev{remind the reader} that our results can be scaled to different pressures and corresponding field strengths.
For example, our simulations 10\,T at 1\,bar should approximately correspond to 1\,T at $0.1 \, \textrm{bar}$ at length and time scales a factor ten larger~\cite{nijdam2020physics}.

\section*{Acknowledgments}
SD is supported by the Science Fund of the Republic of Serbia, Grant No.7749560, Exploring ultra-low global warming potential gases for insulation in high-voltage technology: Experiments and modelling—EGWIn.

\section*{Data availability statement}

\section*{References}
\bibliographystyle{unsrt}
\bibliography{ExB-ref}

\end{document}